# Black hole feedback in the luminous quasar PDS 456

**Authors:** E. Nardini[1]*, J. N. Reeves[1,2], J. Gofford[1,2], F. A. Harrison[3], G. Risaliti[4,5], V. Braito[6], M. T. Costa[1], G. A. Matzeu[1], D. J. Walton[3,7], E. Behar[8], S. E. Boggs[9], F. E. Christensen[10], W. W. Craig[11], C. J. Hailey[12], G. Matt[13], J. M. Miller[14], P. T. O'Brien[15], D. Stern[7], T. J. Turner[16,17], M. J. Ward[18]

**Affiliations:**

[1]Astrophysics Group, School of Physical and Geographical Sciences, Keele University, Keele, Staffordshire ST5 5BG, UK

[2]Center for Space Science and Technology, University of Maryland Baltimore County, 1000 Hilltop Circle, Baltimore, MD 21250, USA

[3]Cahill Center for Astronomy and Astrophysics, California Institute of Technology, Pasadena, CA 91125, USA

[4]INAF – Osservatorio Astrofisico di Arcetri, L.go E. Fermi 5, I-50125 Firenze, Italy

[5]Harvard–Smithsonian Center for Astrophysics, 60 Garden Street, Cambridge, MA 02138, USA

[6]INAF – Osservatorio Astronomico di Brera, Via Bianchi 46, I-23807 Merate (LC), Italy

[7]Jet Propulsion Laboratory, California Institute of Technology, Pasadena, CA 91109, USA

[8]Department of Physics, Technion, Haifa 32000, Israel

[9]Space Science Laboratory, University of California, Berkeley, CA 94720, USA

[10]DTU Space—National Space Institute, Technical University of Denmark, Elektrovej 327, 2800 Lyngby, Denmark

[11]Lawrence Livermore National Laboratory, Livermore, CA 94550, USA

[12]Columbia Astrophysics Laboratory, Columbia University, New York, NY 10027, USA

[13]Dipartimento di Matematica e Fisica, Università degli Studi Roma Tre, Via della Vasca Navale 84, I-00146 Roma, Italy

[14]Department of Astronomy, University of Michigan, Ann Arbor, MI 48109, USA

[15]Department of Physics and Astronomy, University of Leicester, University Road, Leicester LE1 7RH, UK

[16]Physics Department, University of Maryland Baltimore County, 1000 Hilltop Circle, Baltimore, MD 21250, USA

[17]Eureka Scientific Inc., 2452 Delmer Street Suite 100, Oakland, CA 94602, USA

[18]Department of Physics, University of Durham, South Road, Durham DH1 3LE, UK

*Correspondence to: e.nardini@keele.ac.uk

**Abstract**: The evolution of galaxies is connected to the growth of supermassive black holes in their centers. During the quasar phase, a huge luminosity is released as matter falls onto the black hole, and radiation-driven winds can transfer most of this energy back to the host galaxy. Over five different epochs, we detected the signatures of a nearly spherical stream of highly ionized gas in the broadband X-ray spectra of the luminous quasar PDS 456. This persistent wind is expelled at relativistic speeds from the inner accretion disk, and its wide aperture suggests an effective coupling with the ambient gas. The outflow's kinetic power larger than $10^{46}$ ergs per second is enough to provide the feedback required by models of black hole and host galaxy co-evolution.

Disk winds are theoretically expected as a natural consequence of highly efficient accretion onto supermassive black holes (*1*), as the energy radiated in this process might easily exceed the local binding energy of the gas. In the past few years, black hole winds with column densities of $\sim 10^{23}$ cm$^{-2}$ and velocities of $\sim 0.1$ times the speed of light (*c*) have been revealed in a growing number of nearby active galactic nuclei (AGN) through blueshifted X-ray absorption lines (*2*, *3*). Outflows of this kind are commonly believed to affect the dynamical and physical properties of the gas in the host galaxy, and, hence, its star formation history (*4*). However, a complete observational characterization of how this feedback works is still missing. On its own, the detection of narrow, blueshifted features does not convey any information about the opening angle or the ejection site of the wind. This knowledge is critical for measuring the total power carried by the outflow, whose actual influence on galactic scales remains unclear (*5*).

The nearby ($z = 0.184$) radio-quiet quasar PDS 456 is an established Rosetta stone for studying disk winds (*6*–*8*). With a bolometric luminosity $L_{bol} \sim 10^{47}$ erg/s, and a mass of the central black hole on the order of $10^9$ solar masses ($M_{sun}$) (*9*), it is an exceptionally luminous AGN in the local universe and might be regarded as a counterpart of the accreting supermassive black holes during the peak of quasar activity at high redshift. Since the earliest X-ray observations, PDS 456 has regularly exhibited a deep absorption trough at rest-frame energies above 7 keV (*6*), which was occasionally resolved with high statistical significance into a pair of absorption lines at $\sim 9.09$ and 9.64 keV (*7*). Because no strong atomic transitions from cosmically abundant elements correspond to these energies, such lines are most likely associated with resonant K-shell absorption from Fe XXV He$\alpha$ (6.7 keV) and Fe XXVI Ly$\alpha$ (6.97 keV) in a wind with an outflow velocity of $\sim 0.3c$.

The X-ray Multi-Mirror Mission (*XMM*)–*Newton* and Nuclear Spectroscopic Telescope Array (*NuSTAR*) satellites simultaneously observed PDS 456 on four occasions in 2013, between 27 August and 21 September. A fifth observation was performed several months later, on 26 February 2014 (Table S1). The entire campaign caught the quasar in widely different spectral states (Fig. 1). The broad Fe-K absorption trough is conspicuous at all times against a simple baseline continuum, with an equivalent width increasing in absolute value by a factor of $\sim 4$, from 75±45 eV to 350±60 eV (Table S2; uncertainties are given at the 90% confidence level). The large blueshift to $\sim 9$ keV invariably pushes this broad feature close to the edge of the *XMM–Newton* energy band, preventing any evaluation of the intrinsic flux bluewards of the absorption complex. The addition of the *NuSTAR* broadband spectra enabled the accurate determination of the intensity and slope of the high-energy continuum that was lacking before. This unveils a broadened Fe K$\alpha$ emission component, which, combined with the adjacent absorption trough, gives rise to an unmistakable P-Cygni-like profile as produced by an expanding gaseous envelope.

Despite the overall variability, this characteristic outflow signature is clearly detected in each of the five epochs during the course of the campaign (Fig. 2), demonstrating the persistence of a wide-angle accretion disk wind in PDS 456. For a preliminary evaluation of its basic parameters, we first deconvolved the P-Cygni profile with a pair of Gaussian line shapes. For Doppler broadening relative to a uniform velocity field within a radial outflow, the profile width would suggest an aperture of θ ~ 100 degrees (Fig. S1). In reality, this is unlikely to be a simple projection effect, as we might be observing the wind's acceleration region. We therefore applied a custom P-Cygni model to the 2 to 30 keV energy range, only allowing for an absorption-induced continuum curvature (*9, 10*). The success of this attempt implies that the line's profile is indeed compatible with a quasi-spherical (fully covering) outflow with terminal velocity of $0.35\pm0.02c$ (Fig. 3).

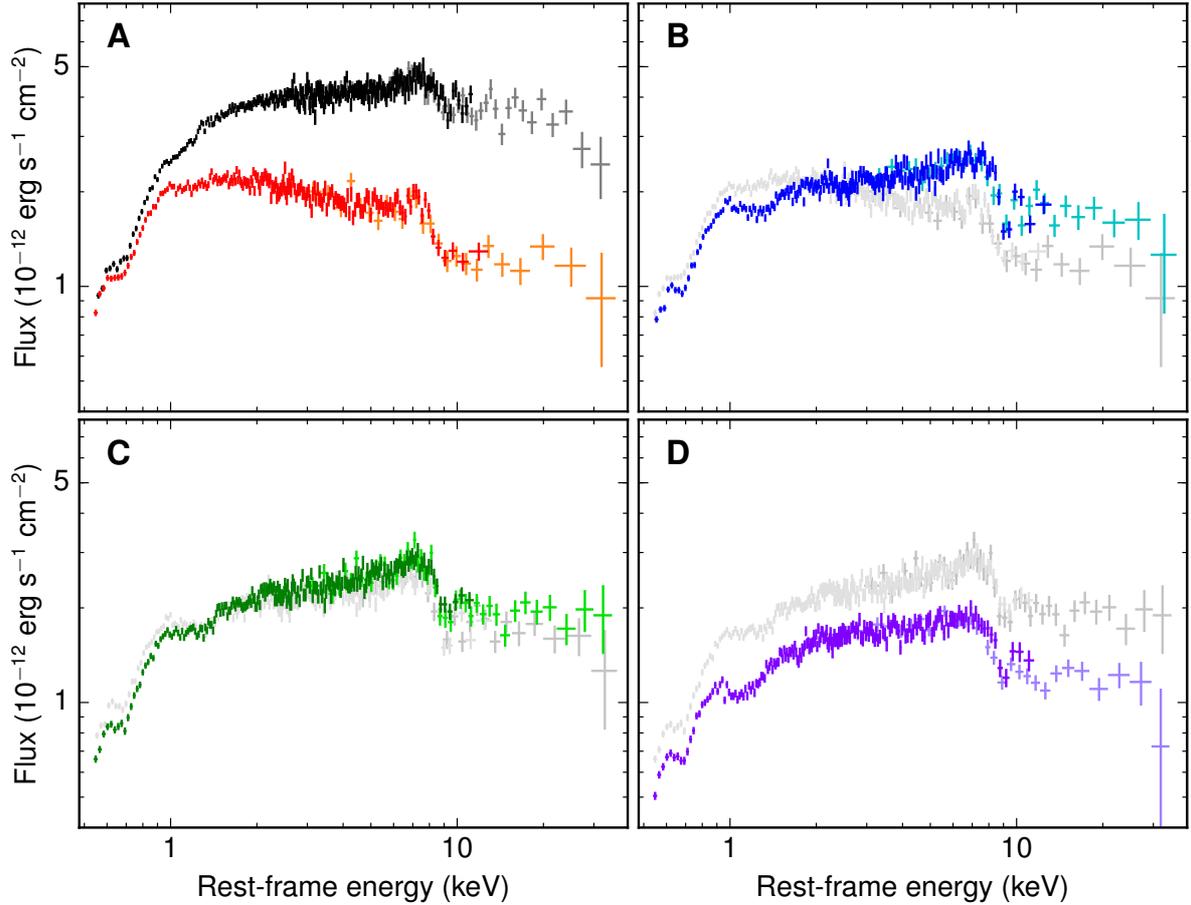

**Fig. 1.** The spectra of the five *XMM–Newton* and *NuSTAR* observations (±1 SD error bars) are plotted in flux density units after factoring out the effects of the instrumental response on the raw photon count rate. All the spectra were re-binned for display purposes only, with the data from the two *NuSTAR* modules merged and shown in a lighter tone. (A) Obs. 1 (black) and Obs. 2 (red). (B) Obs. 2 (shaded) and Obs. 3 (blue). (C) Obs. 3 (shaded) and Obs. 4 (green). (D) Obs. 4 (shaded) and Obs. 5 (purple). Each panel also contains the spectrum of the previous observation in to highlight the extent of the variability.

We subsequently performed a comprehensive spectral and timing analysis, fitting the ~0.5 to 50 keV spectra from all five observations with a model that includes the standard power-law continuum and self-consistent absorption and emission components from photo-ionized gas

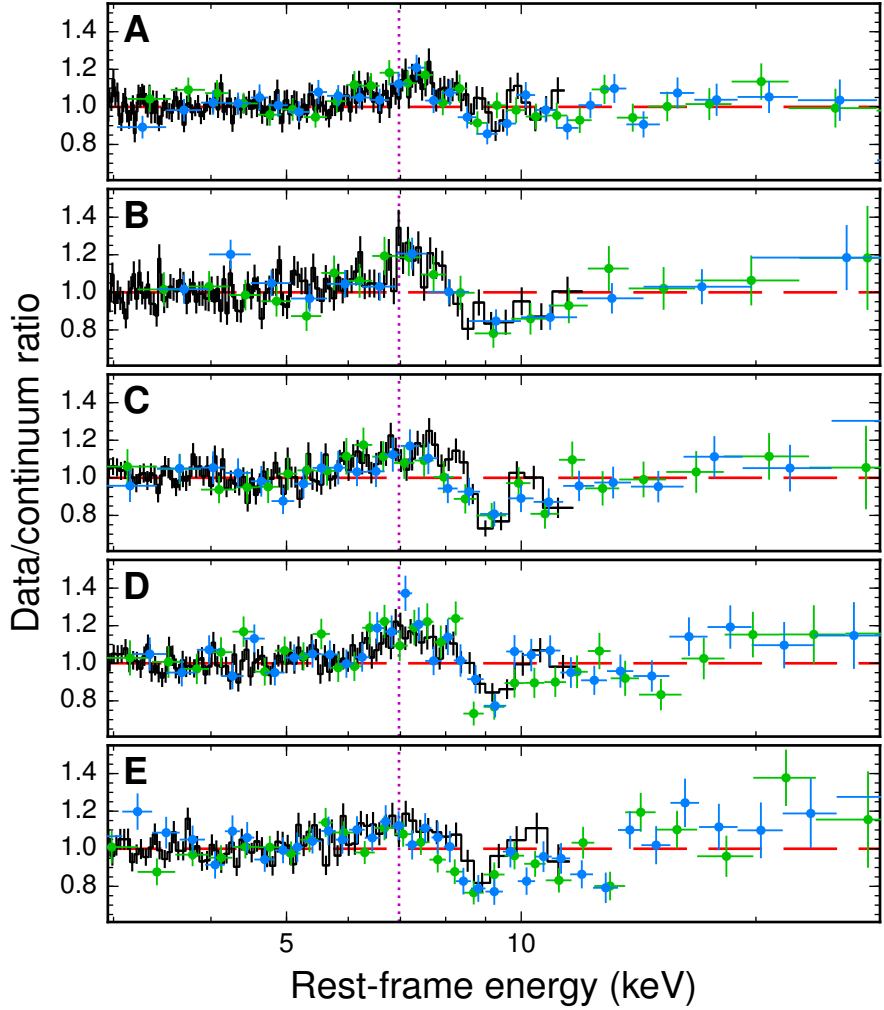

**Fig. 2**. The ratio of the observed emission over the continuum, which was modeled as a partially absorbed power law to reproduce the overall spectral curvature, is shown for *XMM–Newton* data (in black; (±1 SD error bars) and both *NuSTAR* modules (superimposed as green and turquoise dots). The P-Cygni-like profile is evident in each snapshot of the campaign, irrespective of the different flux and spectral states of the source. The peak of Fe Kα emission from the wind lies above 7 keV in each observation, and the absorption trough is centered around 9 keV. The line's profile can be resolved independently at any epoch, with a full width at half-maximum for both components of ~900 eV (or 30,000 km/s at 9 keV). The vertical dotted line marks the rest-frame energy (6.97 keV) of the Fe XXVI Kα transition. (A) Obs. 1. (B) Obs. 2. (C) Obs. 3. (D) Obs. 4. (E) Obs. 5.

(Fig. S2). This model provides a very good description of the data, with no obvious residuals over the whole energy range (Fig. S3). The main physical parameters obtained from the best fit include the hydrogen column density of the highly ionized absorber, $N_\mathrm{H} = 6.9^{+0.6}_{-1.2} \times 10^{23}$ cm$^{-2}$, and its bulk outflow velocity, $v_\mathrm{out} = 0.25 \pm 0.01c$, constant over the five epochs (Table S3). Partial covering by less-ionized gas accounts for most of the continuum spectral variability, which indicates clumpy obscuration within the same stream as observed in lower luminosity active galaxies (*11*).

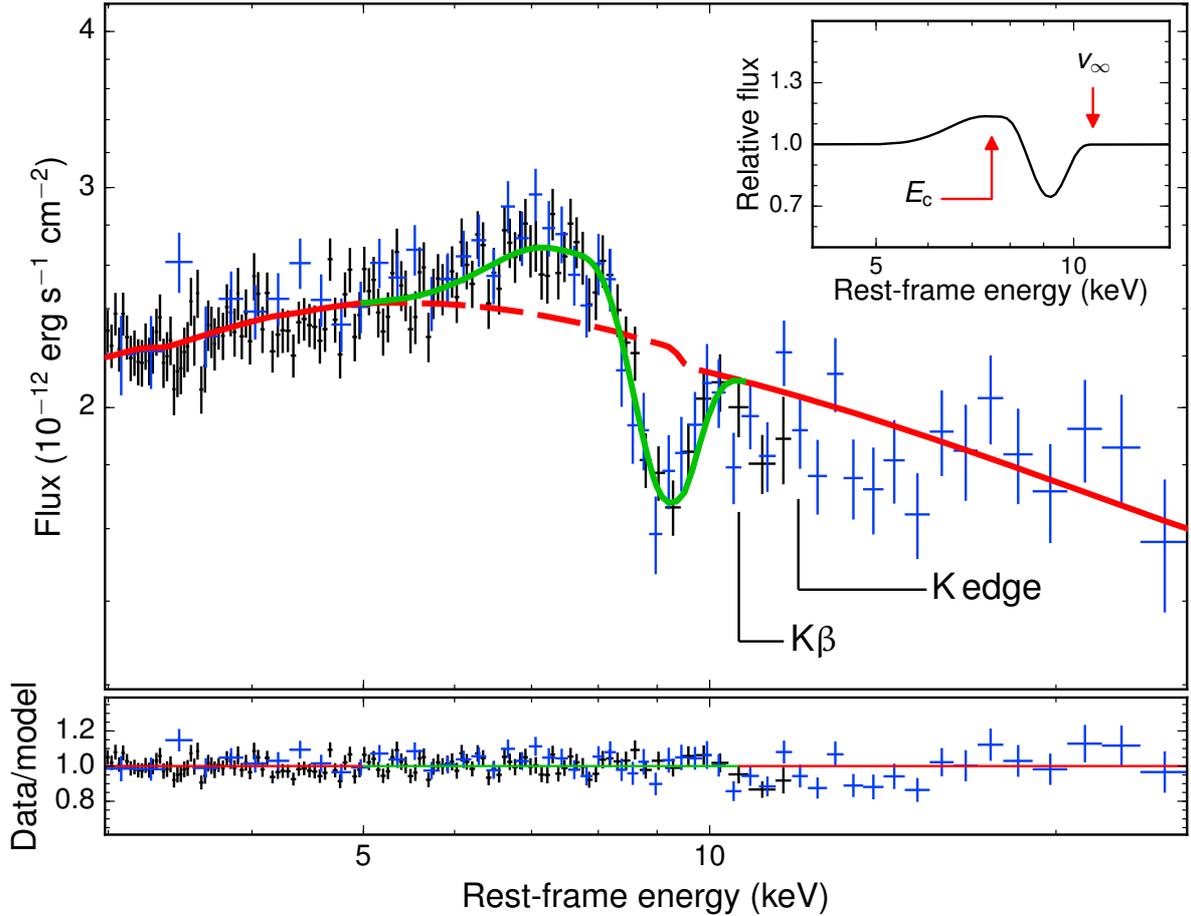

**Fig. 3.** Adopting the same baseline continuum of Fig. 2 (red curve), we fitted the emission and absorption residuals characterizing the Fe-K band by means of a self-consistent P-Cygni profile from a spherically symmetric outflow (green curve). The results are shown for the merged Obs. 3 and Obs. 4, which are separated by only 3 days and are virtually indistinguishable at 2 to 30 keV (Fig. 1). The two *NuSTAR* modules were combined into a single spectrum (plotted in blue; ±1 SD error bars) for display purposes only. The inset contains a graphical explanation of the key parameters of this model: the characteristic energy $E_c$, corresponding to the onset of the absorption component, and the wind terminal velocity $v_\infty = 0.35 \pm 0.02c$, which can be regarded as a measure of the actual outflowing speed of the gas. The bottom panel shows the ratio between the data and the best-fit model. The residual structures above 10 keV are due to the Kβ and K edge absorption features from Fe XXVI. These are not included in the P-Cygni model but are detected with high significance (Table S2), and remove any ambiguity in the identification of the ionic species.

However, the general validity of the disk wind picture is still disputed. It has been proposed that blueshifted absorption might also arise from co-rotating optically thick plasma blanketing the accretion flow, which would be seen in X-rays reflected off the disk surface (*12*). Depending on the exact geometry, the extreme velocities inherent to the inner disk could produce a Fe K-shell feature anywhere between 4 and 10 keV through relativistic Doppler shifts. Previously applied to PG 1211+143, another bright quasar where a similar line complex was revealed (*13*), this scenario calls for a reflection-dominated X-ray spectrum. In PDS 456, this model clearly under-predicts the depth of the 9-keV absorption trough, whose energy and profile allow us to rule out this alternative interpretation (Figures S4 and S5).

Even when blueshifted lines can be safely associated with an outflow, the measure of its mass-loss rate and total energetics is subject to large uncertainties. Remarkably, these observations supply a robust estimate of the solid angle $\Omega$ filled by the wind, which is computed from the amount of absorbed ionizing radiation that is re-emitted across the spectrum. The average value is $\Omega = 3.2 \pm 0.6\pi$ sr, and each individual observation is consistent with $\Omega$ exceeding $2\pi$ sr (*9*). No explicit information on $\Omega$ for individual sources existed previously, because the absorption components detected in both ultraviolet and X-ray spectra only probe the line of sight. Consequently, this quantity has thus far been assumed rather than directly determined from the data – for example, following the occurrence frequency of outflows amongst local AGN (*14*), or through an a priori selection of a reasonable wind geometry (*15*).

Besides the degree of collimation, critical information is required on the starting point of the wind ($R_{in}$), for which only indirect arguments are usually available (*16*). In PDS 456, the gas location is directly constrained by changes of the wind emission intensity in response to the hard X-ray continuum. The most significant variation, when the Fe K line follows a decrease by a factor of ~3 of the 7 to 30 keV flux, takes place between the first two observations, separated by ~10 days. The corresponding light-crossing time translates into a maximum distance of a few hundreds of gravitational radii ($r_g = GM/c^2$, where $G$ is the gravitational constant and $M$ is the black hole's mass) from the illuminating source. This spatial extent is consistent with the degree of ionization of the gas, and with predictions of hydro-dynamical models of radiation-driven accretion disk winds (*17*). An independent corroboration of the above estimate comes from the variability timescale of ~1 week of the Fe-K absorption feature, probed during the monitoring of PDS 456 by the *Suzaku* X-ray satellite in early 2013 (*18*). Indeed, the historical behavior of the K-shell absorption line(s) appears to be in keeping with a persistent wind where the gas is in photo-ionization equilibrium with the local radiation field.

We therefore adopt $R_{in} = 100\ r_g \sim 1,000$ astronomical units, and take conservative values for the other physical and geometrical quantities involved (supplementary online text). With all the relevant pieces of information now available, we determine a mass outflow rate at the base of the wind of $\dot{M}_{out} \sim 10\ M_{sun}$/year, corresponding (for a mass to radiation conversion efficiency $\eta \sim 0.1$) to about half of the Eddington accretion rate, the limit at which gravitational infall is balanced by outward radiation pressure. Consequently, the kinetic power of the wind is $P_{kin} \sim 2 \times 10^{46}$ erg/s, or 20% of the bolometric luminosity of the quasar.

According to models and simulations (*19*, *20*), the deposition of a few percent of the total radiated energy is sufficient to prompt an appreciable feedback on the host galaxy. As most of the kinetic output will be ultimately transferred to the surrounding gas (*21*), these conditions are met even if $R_{in}$ is set to the escape radius (~30 $r_g$ for $v_{out} = 0.25c$), the minimum plausible starting point of the wind.

The mechanical energy released over a period of $10^7$ years (i.e. one tenth of a typical quasar lifetime) is close to $10^{61}$ erg, comparable to or exceeding the expected binding energy of the galactic bulge in a system like PDS 456. The large opening angle evinced here suggests that the coupling of the outflow with the gas in the host galaxy will be effective, as required for strong negative feedback. We are then possibly witnessing the initial stage of the sweeping process that leads to molecular mass losses of hundreds to thousands $M_{sun}$/year even in sources with no powerful radio jets (*22*). For a commensurate kinetic luminosity, in fact, the observed galaxy-wide outflows entail considerable mass loading and momentum boost.

A fundamental correlation (the so-called $M$–$\sigma_*$ scaling relation) exists between the mass of the central black holes and the stellar velocity dispersion of galactic bulges on kpc scales (*23, 24*) – that is, hundreds of times beyond the gravitational sphere of influence of the black holes themselves. It is still debated whether this is the product of a feedback-driven black hole/host galaxy co-evolution (*25*), or of a hierarchical assembly through galaxy mergers (*26*). Identifying wide-angle accretion disk winds in the Eddington-limited regime lends weight to the idea that AGN have a substantial impact on the surrounding environment. As a rare, nearby analogue of the luminous AGN population at high redshift, PDS 456 shows a flavor of cosmic feedback, believed to have operated at the peak of the quasar epoch about 10 billion years ago. In distant galaxies in a similar activity phase, such powerful winds would have provided the energy and momentum to self-regulate the black hole growth and control the star formation in stellar bulges, leaving the present-day scaling relations as a record of this process (*27*).

**Acknowledgments:**

This research was supported under the U.K. Science and Technology Facilities Council grant ST/J001384/1, and is based on X-ray observations obtained with the *XMM–Newton* and *NuSTAR* satellites. *XMM–Newton* is an European Space Agency (ESA) science mission with instruments and contributions directly funded by ESA member states and the National Aeronautics and Space Administration. The *NuSTAR* mission is a project led by the California Institute of Technology, managed by the Jet Propulsion Laboratory, and funded by NASA. We thank the *NuSTAR* Operations, Software, and Calibration teams for support with execution and analysis of these observations. We also acknowledge financial support from the Italian Space Agency under grant ASI-INAF I/037/12/0 (G.R. and G.M.); the Italian National Institute for Astrophysics under grant PRIN-INAF 2012 (G.R.); the I-CORE program of the Planning and Budgeting Committee, the Israel Science Foundation under grants 1937/12 and 1163/10, Israel's Ministry of Science and Technology (E.B.); and NASA under grants NNX11AJ57G and NNG08FD60C (T.J.T.). The data are stored in the science archives of the two X-ray observatories involved, and will become publicly available on 25 March 2015 (*XMM–Newton*) and with the upcoming DR6 data release (*NuSTAR*).

**Supplementary Materials:**

Materials and Methods

Supplementary Text

Figures S1-S5

Tables S1-S3

References (*28–63*)

**Supplementary Materials:**

**Materials and Methods**

<u>1. Observations and data reduction</u>

PDS 456 was observed simultaneously with the *XMM–Newton* (*28*) and *NuSTAR* (*29*) X-ray satellites four times between August 27 and September 21, 2013 (hereafter Obs. 1–4), and once again on February 26–27, 2014 (Obs. 5). The details of each observation are summarized in Table S1.

The *XMM–Newton* data were reduced and calibrated according to the standard procedures, using the Scientific Analysis System version 13.0 and the latest version of the calibration database made available by the *XMM–Newton* Science Operations Centre. The observations were taken in the Large Window mode of the European Photon Imaging Camera (EPIC) pn CCD detector, with the medium optical blocking filter in place. We checked each light curve for particle background flares, removing any time intervals above the threshold of 0.5 count/s in Obs. 1–4, and 1 count/s in Obs. 5. The resultant cumulative net exposure was 483 ks. Source and background spectra were extracted from circular regions of 35" and 60" radius, respectively. The net count rates over the 0.45–10 keV energy band are in the range ~1.3 (Obs. 5) to 3.1 counts/s (Obs. 1). At these levels, the effects of photon pile-up on the spectra are negligible.

Although the *XMM–Newton* analysis was eventually performed on pn spectra only, we carried out a self-consistency check with the data from the two MOS detectors, whose effective areas drop quickly at the energies of the Fe-K features. A remarkable agreement was found in each observation down to ~0.45 keV, which was therefore assumed as the lower energy bound in our fits. The spectra were grouped to a minimum of 100 counts per energy bin in order to allow the use of $\chi^2$ minimization during the spectral analysis, which was performed with the XSPEC fitting package version 12.8.1g.

We also considered the source photometry in the six wide-band *V*, *B*, *U*, *UVW1*, *UVM2* and *UVW2* filters of the *XMM–Newton* Optical Monitor (OM). We averaged the count rates from the small high-resolution window at the center of the OM field of view over the different observations, with a 10% uncertainty added in quadrature to account for systematic errors in the flux of the calibration stars. We applied an absorption correction based on the standard extinction law (*30*), with values of *A*(*V*) and *E*(*B–V*) appropriate for the source (*31, 32*). The five OM data sets were used to determine the shape of the optical–ultraviolet (UV) ionizing continuum.

The simultaneous *NuSTAR* level 1 data products were processed with the *NuSTAR* Data Analysis Software package version 1.3.1 included in the 6.15.1 HEASoft release. Event files were produced, calibrated, and cleaned using the standard filtering criteria and the latest files available in the *NuSTAR* calibration database (CALDB version 20131223). The total net exposure, after the exclusion of periods of source occultation by Earth and passages on the South Atlantic Anomaly, was around 299 ks for both *NuSTAR* co-aligned telescopes, with corresponding Focal Plane Modules A (FPMA) and B (FPMB). Extraction radii were 80" for both source and background spectra, which were grouped to have at least 50 counts per bin.

<u>2. Parameterization of the P-Cygni profile</u>

The considerable curvature characterizing PDS 456 in the soft X-rays and the spectral variability on timescales as short as weeks (Fig. 1) have been usually attributed to reprocessing in multiple layers of absorbing gas along the line of sight. In the wake of this, we performed a preliminary inspection of the hard X-ray spectra between 2–30 keV using a partially covered continuum model. We jointly fitted all the simultaneous *XMM–Newton* and

*NuSTAR* observations in the 2–10 and 3–30 keV energy intervals, respectively. Whilst not necessarily physically motivated at this stage, a good description of the overall continuum shape was obtained through a power law with photon index $\Gamma \sim 2.3$, partially covered (covering fraction $f_{cov} \sim 0.1$–$0.4$) by a neutral gas column of $N_H \sim 10^{23}$ cm$^{-2}$ (Table S2). Figure 2 shows the five ratio plots of the data against this model.

The P-Cygni-like feature in the Fe-K spectral region can be fitted with two Gaussian profiles (one in emission and the other in absorption) at $E_{rest} = 6.97$ keV, corresponding to the rest-frame energy of the Fe XXVI Ly$\alpha$ resonance line. Additional residuals remain beyond ~9 keV, indicating the possible presence of further atomic structures associated with the absorber. To investigate whether all these hard X-ray features arise from the same stream of outflowing gas, we included another Gaussian absorption line and a photoelectric edge into the baseline continuum model. The former was identified with the Ly$\beta$ transition at $E_{rest} = 8.25$ keV, while the latter was introduced to account for the K-shell absorption edge at $E_{rest} = 9.28$ keV (the minimum photon energy capable of stripping the last electron from Fe XXVI). We fixed all the rest-frame energies to their expected laboratory values, assuming a common broadening $\sigma_K$ and allowing the relative outflow velocity between the emission line and the absorption components to vary.

Such an ensemble offers an excellent description of the Fe-K spectral range in each observation (Table S2). If attributed to Fe XXVI Ly$\alpha$, the emission profile appears to be slightly blueshifted with respect to the systemic velocity of PDS 456, with $v_{out} = 0.029\pm0.017c$ (uncertainties are given at the 90% confidence level). On the other hand, a substantially larger outflow velocity $v_{out} \sim 0.24c$ is implied for the Fe XXVI Ly$\alpha$, Ly$\beta$ and K-shell edge absorption series, whose simultaneous and self-consistent detection rules out any identification with lower ionization species (e.g. Fe XXV). Both the emission and absorption profiles are well resolved, with $\sigma_K = 380^{+50}_{-30}$ eV. This corresponds to a velocity dispersion of ~13,000 km/s across the flow (~30,000 km/s full width at half-maximum).

This model returns a very good fit of the entire 2–30 keV spectra ($\chi^2 = 2540.8$ for 2582 degrees of freedom). The equivalent width (EW) of the Ly$\alpha$ and Ly$\beta$ absorption lines follows a general trend of being weaker at the start of the campaign and becoming systematically larger afterwards, while the overall count rate decreases accordingly. The Ly$\alpha$ EW varies from $-75\pm45$ eV (in Obs. 1) to $-350\pm60$ eV (in Obs. 3), and the Ly$\beta$ exhibits a similar behavior (Table S2). The maximum optical depth of the absorption edge ($\tau_{max}$) is in keeping with the observed EW trend, i.e. $\tau_{max}$ is higher when the Ly$\alpha$ and Ly$\beta$ EWs are larger. This is consistent with a line variability driven by changes in the opacity of the absorbing gas.

The apparent Fe XXVI P-Cygni-like feature hints at a large covering factor of highly ionized matter in proximity of the central black hole. The Sobolev approximation with exact integration (SEI) is a commonly used method to calculate the P-Cygni line profiles associated with the expansion of spherically symmetric stellar winds (*33*). A SEI-based model has been also developed to specifically reproduce Fe-K absorption features in AGN (*10*). We assumed a velocity field of the form $w = w_0 + (1-w_0)(1-1/x)^\gamma$, where $w = v/v_\infty$ is the wind velocity normalized to the terminal velocity, $w_0$ is the dimensionless velocity at the wind photosphere, $x = r/R_0$ is the radial distance in units of the photospheric radius $R_0$, and $\gamma$ governs how the velocity scales with distance. Since the exact values of $\gamma$ and $w_0$ have little effect on the derived P-Cygni profiles, we chose $\gamma = 1$ and $w_0 = 0.01$. For this given velocity field, the line optical depth varies as $\tau(w) \propto \tau_{tot} w^{\alpha_1}(1-w)^{\alpha_2}$, where the quantities $\alpha_1$ and $\alpha_2$ determine the smoothness of the resultant P-Cygni profile, with larger values leading to more gradual shapes (*34*).

We then fitted again all five observations simultaneously between 2–30 keV, replacing the pair of Kα Gaussian lines with a self-consistent P-Cygni profile. For simplicity, we assumed $\alpha_1 = \alpha_2$ (in practice they cannot be constrained individually), and fixed the underlying continuum slope to the best-fit value from the previous model, to prevent any degeneracy between the spectral curvature and the P-Cygni parameters. The fit variables include the characteristic energy of the profile ($E_c$), corresponding to the onset of the absorption component, the terminal velocity of the wind ($v_\infty$), and the total optical depth of the gas ($\tau_{tot}$). While $E_c$ and $v_\infty$ were required to be the same over the five sequences, $\tau_{tot}$ was left free to vary independently to account for changes in the line intensity, and increases from $\tau_{tot} = 0.05\pm0.01$ (Obs. 1) to $\tau_{tot} \sim 0.10$–$0.15$ (Obs. 2–5), confirming that the opacity of the reprocessing gas is lower at the beginning of the campaign (Table S2). The characteristic energy of the P-Cygni line is $E_c \sim 7.4\pm0.1$ keV (rest-frame), and its terminal velocity is $v_\infty = 0.35\pm0.02c$. None of the higher order features are included in this model by default. Once the Kβ absorption line and the K-shell edge are re-introduced, the fit statistic is $\chi^2/\nu = 2613.1/2604$, comparable to that achieved in the Gaussian case.

3. Broadband modeling

The P-Cygni fit implies that the disk wind in PDS 456 is qualitatively consistent with a quasi-spherical flow. Our next goal is to quantify both the properties of the absorbing gas and the solid angle of the wind, by using the XSTAR photo-ionization code (*35*). The ionization balance is sensitive to the distribution of ionizing photons between 1–1,000 Rydberg (i.e., from 13.6 eV to 13.6 keV), as determined by the shape of the underlying UV to X-ray radiation field. By combining the EPIC pn and *NuSTAR* X-ray spectra with the data from the six *XMM–Newton* OM photometric bands, we thus obtained a proper input ionizing continuum for the generation of the relevant XSTAR absorption and emission models. The spectral energy distribution (SED) can be approximated from 2 eV to 30 keV by a three-segment broken power law. We adopted an X-ray photon index $\Gamma = 2.4$ to match the mean hard X-ray spectral slope found by *XMM–Newton* and *NuSTAR*, while the time-averaged OM data follow a $\Gamma = -0.7$ trend. The connecting UV to X-ray slope at 10–500 eV is $\Gamma \sim 3.3$. This model predicts a total ionizing luminosity $L_{ion}$ of the order of $5 \times 10^{46}$ erg/s in the 1–1,000 Rydberg bandpass. We also investigated a more realistic scenario, replacing the broken power law with multicolor blackbody emission from an accretion disk with photon temperature $kT \sim 5$–$10$ eV. This is similar to what is expected for a black hole with mass of $\sim 10^9 M_{sun}$ accreting at nearly Eddington rate (*36*), and entails $L_{ion} \sim 10^{47}$ erg/s, as is likely the case for PDS 456 (see below).

The XSTAR code computes the radiative transfer through a spherically symmetric shell of absorbing gas, yielding two grids of absorption and emission spectra. Absorption tables have three main free parameters: the column density $N_H$ of the gas, its ionization parameter $\xi$ (defined as $\xi = L_{ion}/nR^2$, where $n$ is the gas number density, and $R$ the distance from the ionizing source), and the absorber's redshift $z$. Emission tables also take a normalization, which depends on the solid angle $\Omega$ subtended by the reprocessing gas shell to the illuminating source. With the information gathered from the SED analysis, we generated a pair of absorption/emission XSTAR grids with Solar abundances suitable for PDS 456, and covering the $5 \times 10^{21} < N_H/\mathrm{cm}^{-2} < 2 \times 10^{24}$ and $1 < \log(\xi/\mathrm{erg\ cm\ s}^{-1}) < 8$ parameter ranges. These components were used to model all of the effects imparted by the accretion disk wind on the observed spectra. Based on the observed width of the Fe-K features, we imposed a velocity smearing $v_K = 15{,}000$ km/s. Another grid with a modest broadening of 100 km/s, which is more typical of the soft X-ray warm absorbers on parsec scales (*37, 38*), was produced to account for a more distant layer of absorbing gas.

The ultimate interpretation of the observed spectral variability in the broadband X-ray emission of PDS 456 involves a combination of changes in the properties of the absorbing gas and subtle variations of the primary continuum. The mathematical structure of the model we applied can be represented as follows:

$$galabs \times warmabs \times windabs \times (pcabs \times powerlaw + windem + softexc). \quad (1)$$

The various components, identified above through self-evident tags, are illustrated in Fig. S2. The global fit, performed simultaneously over all the data sets, yields a statistic of $\chi^2 = 4372.5$ for 4180 degrees of freedom (while $\chi^2/\nu = 2651.3/2610$ above 2 keV, for a null hypothesis probability of 0.28) with no conspicuous residuals (Fig. S3). Below we proceed to a more detailed discussion of the model components.

- *galabs:* Galactic absorption was included in all our fits, with standard cross sections and Solar abundances (*39*). The hydrogen column density was fixed throughout to $N_{H,Gal} = 2.0 \times 10^{21}$ cm$^{-2}$, as appropriate for the coordinates of the source (*40*).

- *powerlaw:* The primary continuum was modeled using a power law, representing the intrinsic X-ray emission of PDS 456 before reprocessing. In order to fit the broadband spectra, some intrinsic variability is required, with the photon index $\Gamma$ lying in the range ~2.3–2.7 and a factor of ~3 difference in the 7–30 keV flux across the observations (Table S3).

- *windabs:* Blueshifted K-shell absorption from highly ionized iron was reproduced by means of an XSTAR grid. Any changes in the depth of Fe-K absorption can be solely ascribed to ionization changes, with log ($\xi$/erg cm s$^{-1}$) decreasing from $6.30^{+0.10}_{-0.16}$ in Obs. 1, when the feature is weakest, to a minimum of $5.75^{+0.06}_{-0.16}$ in Obs. 3, when it is strongest. Qualitatively, these variations trace the behavior of the continuum flux. Conversely, the column density is consistent within the errors over the entire campaign, and was therefore tied across the five data sets returning a best-fit value of $N_H = 6.9^{+0.6}_{-1.2} \times 10^{23}$ cm$^{-2}$. The blueshift of the high-$\xi$ absorber turns out to be virtually constant (Table S3), implying an outflow velocity $v_{out} = 0.25\pm0.01c$, in agreement with the preliminary Gaussian-line fit. Moreover, the trend of decreasing $\xi$ (hence increasing opacity) confirms the findings of the P-Cygni analysis.

- *windem:* Emission associated with the disk wind was included in the model as an XSTAR grid. Indeed, photons absorbed within a quasi-spherical or wide-angle wind are subject to non-negligible re-emission along the line of sight, thereby leading to a P-Cygni-like profile. For this component, we assumed the same ionizing SED as per the absorption tables, with equal velocity smearing, ionization state and column density. The only free parameters of the emitter are its normalization and its redshift (common to the five epochs), which prefers a moderate outflow speed $v_{out} = 0.012\pm0.010c$.

- *pcabs:* Partial covering absorption has been established by previous studies as the origin of the pronounced spectral curvature below ~3–4 keV in PDS 456 (*6–8*). For consistency, we accounted for this component by using the same custom XSTAR absorption table generated for the high-ionization wind. For a best-fit ionization parameter of log ($\xi$/erg cm s$^{-1}$) = 2.9, we found the overall spectral shape to be well explained by a variable column density, decreasing from $N_H \sim 2.5 \times 10^{23}$ cm$^{-2}$ in Obs. 1–2 to $N_H \sim 0.7 \times 10^{23}$ cm$^{-2}$ in Obs. 5, with small changes of the covering fraction in the range $f_{cov} \sim 0.3–0.5$ (Table S3). Most interestingly, the partially covering gas accepts some degree of blueshift relative to the local frame of PDS 456, with an outflow velocity commensurate with that of the Fe-K absorber ($v_{out} \sim 0.25c$). The fit is substantially worse if this component is forced to the systemic velocity of PDS 456 ($z = 0.184$), with $\Delta\chi^2 \sim +80$ for the same degrees of freedom. Given the lower ionization of this gas, and the slow variability over several weeks, it is then possible that partial covering arises from more distant clumps within the wind, as found in other AGN (*11*).

• *softexc:* Similarly to what had been previously reported (*6, 41*), the data seem to require a soft excess below ~1 keV. As the origin of this component (ubiquitous among AGN) is still largely ambiguous, here we modeled it through an ad hoc broad Gaussian emission line. Indeed, the exact parameterization of the soft excess does not affect in any way the fit of the high-ionization Fe-K features, nor our conclusions on the properties of the wind in PDS 456.

• *warmabs:* A fully covering absorber of low ionization log ($\xi$/erg cm s$^{-1}$) = 0.31±0.02 and minor velocity broadening, possibly representing a distant warm absorber, also contributes to the observed spectral curvature. Its column $N_H$ = 4.3$^{+0.2}_{-0.1}$ × 10$^{21}$ cm$^{-2}$ is essentially constant in Obs. 1–4, and appears to be lower in Obs. 5, reaching $N_H$ = 1.8±0.2 × 10$^{21}$ cm$^{-2}$. This suggests that the typical variability timescale for this component is a few months long. Due to the low column density value, this putative warm absorber is only manifest below ~2 keV, having no effect on the Fe-K region of the spectrum and the derived physical parameters.

**Supplementary Text**

1. Alternative models

The detection of blueshifted absorption lines in the hard X-ray spectra of AGN is usually associated with fast outflows. However, this interpretation is not unequivocal. It has also been proposed that some of these features could instead arise from resonant absorption in a layer of highly ionized, optically thick plasma located above the accretion disk, and co-rotating with it (*42*). The main advantage of this scenario (referred to as 'disk absorption' in the following for ease of discussion) is that relativistic velocities are natural in proximity of the central black hole, and there is no need to invoke any wind launching mechanism. Here absorption is imprinted on the X-rays reflected off the disk surface, while any direct power-law continuum would be left unaffected (see Fig. 1 in (*42*) for a geometrical illustration). Thus, for prominent absorption lines to be produced, the observed 2–10 keV spectrum has to be dominated by the reflected component, rather than by the primary emission. In other words, the reflection strength RS (*43*) has to be much larger than one. In this framework, the Fe XXVI K$\alpha$ feature can be shifted from its rest-frame energy of 6.97 keV up to 9 keV and possibly beyond, depending on the exact geometry. This model has been successfully applied (*12*) to reproduce the absorption line observed in PG 1211+143, whose blueshift would correspond to an outflow velocity of ~0.10–0.15$c$ in the wind scenario. Fe-K emission from the disk, once blurred by relativistic effects, accounts for the remains of the claimed P-Cygni profile, which, however, emerged only after combining multiple observations taken several years apart (*13*).

We have tested this picture for PDS 456 on Obs. 3, during which the equivalent width of the absorption line appears to be larger. Energies below 3 keV were initially neglected to avoid the complications due to additional soft X-ray absorption and emission components. To mimic the presence of a hot atmosphere blanketing the disk, we defined a model with the following structure:

$$galabs \times (powerlaw + relblur \times diskabs \times diskref).$$

The standard power-law continuum also provides the input illumination for the disk reflection component (*44*), which is convolved with a blurring kernel (tagged as *relblur* above) that imparts the required relativistic distortions (*45*). The 9-keV absorption is included as an XSTAR table (*diskabs*), linked to the reflected spectrum only and subject to the same smearing. When varying freely, this model tries to account for the flux drop at the onset of the absorption trough as either the blue wing of a blurred Fe Kα emission line or a K-shell edge, rejecting any effect from the hot-gas absorber through either a vanishing column density or an extremely high ionization. This pulls down the continuum, leaving obvious residuals at the upper end of the *XMM–Newton* bandpass, while at higher energies the shape of the *NuSTAR* spectra is gradually recovered by a modest Compton reflection hump (*46*).

For illustration purposes, we fixed the values of the dimensionless spin parameter $a^* = cJ/GM^2 = 0.7$ (*J* and *M* are the black hole's angular momentum and mass) and of the disk inclination $i = 75$ degrees, as dictated by the desired blueshift of the Fe XXVI Kα line (*42*). We also set the ionization of the absorbing layer to log ($\xi$/erg cm s$^{-1}$) = 6. Figure S4 shows the fit residuals of this disk absorption model against those of the broadband wind model. The 9-keV feature is clearly missed, even if 10× Solar Fe abundance is adopted. It is in fact the small reflection fraction of RS ~ 0.6 that prevents the absorption profile from reaching an adequate depth. It is still possible that more complex geometrical configurations than the one considered here can reproduce the Fe XXVI Kα feature. Its apparent consistency over the different epochs, however, requires all the X-ray spectral states exhibited by PDS 456 in the course of this campaign to be reflection-dominated. This seems quite unlikely, given the remarkable changes in both intensity and overall spectral shape.

In order to investigate this possibility, and to assess the reflection contribution in emission, we re-fitted the broadband 0.5–30 keV spectra of all observations after including in the wind model a self-consistent blurred reflection component originated by a lamppost source located above the disk (*47*). Any parsec-scale reflector can be disregarded due to the lack of a narrow Fe K line, while an independent soft excess component (again parameterized as a broad Gaussian line) is confirmed to be highly significant. The improvement achieved through this model is questionable ($\Delta\chi^2 \sim -27$ for eight degrees of freedom less), and disk reflection is not statistically required in four out of five observations (Obs. 1–4) according to an *F*-test. Even in Obs. 5, the height of the illuminating source is several tens of gravitational radii; this is not compatible with an extreme light bending regime and the consequent suppression of the direct continuum due to the focusing of the radiation towards the disk (*48*). Incidentally, the best-fit statistics cannot be recovered by replacing either the partial covering or the wind emission with disk reflection. Both cases were checked separately, and revealed a tendency to overestimate the *NuSTAR* spectra, which do not show any evidence of a reflection hump at ~20 keV (as opposed to many other local AGN). This is also clear from the comparison between Obs. 2 and the 2007 *Suzaku* observation (Fig. S5), where a hard excess indicative of a Compton hump was tentatively detected (*7, 49*).

In conclusion, we can definitely rule out the association of the 9-keV absorption trough in PDS 456 with a non-wind scenario, while the presence of a reflection component (either off the disk or within the wind itself) is not imposed by the data, and has been neglected.

2. Mass outflow rate

Let $\rho(r)$ and $v(r)$ be the local mass density and velocity of the wind, respectively. The mass outflow rate through a thin, spherical shell of radius *r* and solid angle $\Omega$ is given by $\dot{M}_{\text{out}} = \Omega r^2 \rho(r) v(r)$. For cosmic elemental abundances and full ionization $\rho(r) \sim 1.2 m_p n(r)$, where $m_p$ is the proton mass and $n(r)$ is the ion number density. Neglecting the constant factor of order unity, and assuming that the wind has a constant velocity $v(r) = v_{\text{out}}$, we obtain $\dot{M}_{\text{out}} \sim \Omega r^2 m_p n(r) v_{\text{out}}$. The product $r^2 n(r)$ is constant, and can be associated to observables as follows:

$$N_{\text{H}} = \int_{R_{\text{in}}}^{\infty} n(r)\,\text{d}r \sim \int_{R_{\text{in}}}^{\infty} \left(\frac{\dot{M}_{\text{out}}}{\Omega m_{\text{p}} v_{\text{out}}}\right) r^{-2}\,\text{d}r = \frac{\dot{M}_{\text{out}}}{\Omega m_{\text{p}} v_{\text{out}} R_{\text{in}}}, \qquad (2)$$

where the inner radius $R_{\text{in}}$ is the starting point of the wind. The final expression brings out all the relevant dependencies for the mass outflow rate:

$$\dot{M}_{\text{out}} \sim \Omega N_{\text{H}} m_{\text{p}} v_{\text{out}} R_{\text{in}}. \qquad (3)$$

We now discuss the measures of all the physical quantities involved.

2.1 Solid angle

Hydrodynamical models of the radiation transport through ionized biconical winds have been calculated to explain the P-Cygni profile of strong UV lines in bright Galactic sources such as cataclysmic variables (*50, 51*). The current quality and resolution of the X-ray spectra from AGN only allows a qualitative comparison of the P-Cygni feature detected in PDS 456 with the atlases of synthetic P-Cygni profiles. Given the depth of the absorption trough, the line of sight has to lie inside the wind. Assuming a symmetric (Gaussian) shape and Doppler broadening, the continuous line profile maps the components of the velocity field projected along the direction to the observer (Fig. S1). From the red wing of the emission and the central energy of the absorption, we infer a line-of-sight velocity range of $-0.04c$ to $0.25c$, which implies an opening angle for a radial outflow of $\theta \sim \cos^{-1}(v_{\text{min}}/v_{\text{max}}) \sim 100$ degrees, corresponding to a solid angle $\Omega \sim 3\pi$. While any wind acceleration is neglected, this value is consistent with the outcome of the P-Cygni fit.

The two main arguments to evaluate the solid angle occupied by the wind are actually independent on both the exact geometrical structure and the local velocity field, as outlined below:

• The normalization $\kappa_{\text{obs}}$ of the XSTAR wind emission table (*windem*) in the broadband analysis includes direct information on the fraction of solid angle covered by the reprocessing gas shell. The grids were generated assuming $\Omega = 4\pi$ and a 1–1,000 Rydberg luminosity $L_{\text{ion}} = 10^{47}$ erg/s. This provides a reference value against which the normalizations delivered by the best fit (which depend on the actual covering fraction and ionizing luminosity) can be interpreted. Following the adopted SED shape, we first rendered explicit the relation between $L_{\text{ion}}$ and the 7–30 keV luminosity, which is a measurable quantity and refers to most critical energy range (i.e. above the Fe XXVI K$\alpha$ threshold) given the ionization state of the gas. We can then assess the solid angle through the following expression:

$$\frac{\Omega}{4\pi} = 3.15 \times 10^3 \kappa_{\text{obs}} \left(\frac{L_{7-30\,\text{keV}}}{10^{44}\,\text{erg/s}}\right)^{-1}. \qquad (4)$$

Note that, in terms of the spectral analysis, $\kappa_{\text{obs}}$ is ultimately driven by the EW of the Fe XXVI K$\alpha$ emission line (Fig. S2). The resulting ratios for the five observations 1–5 are ~0.83 (> 0.49), 1.02 (> 0.59), 0.42±0.18, $0.80^{+0.16}_{-0.20}$, 0.99 (> 0.75). The weighted average is 0.79±0.15, thus $\Omega = 3.2 \pm 0.6\pi$.

• Alternatively, we can compute $\Omega/4\pi$ as the fraction of the absorbed continuum luminosity that is re-emitted within the wind, $L_{\text{em}}/L_{\text{abs}}$. We derived these values over the 0.5–30 keV range, thanks to the simultaneous, broadband spectral coverage. Since both $L_{\text{em}}$ and $L_{\text{abs}}$ depend in a complex way from several parameters ($\Gamma$, $N_{\text{H}}$, $\xi$ and flux normalizations), the errors were estimated through a bootstrap re-sampling technique over the parameters involved. As the distributions are not perfectly symmetric, the uncertainties correspond to the 16[th] and 84[th] percentiles, and are comparable to the 1.645 SD (90 per cent confidence level) error bars in the Gaussian approximation. We obtain the following measures of $\Omega$ for the individual observations: $1.9\pm0.7\pi$ (Obs. 1), $1.4\pm0.6\pi$ (Obs. 2), $0.7^{+0.3}_{-0.2}\pi$ (Obs. 3), $1.6\pm0.4\pi$ (Obs. 4), $1.8^{+0.6}_{-0.4}\pi$ (Obs. 5). The weighted value against the average absorbed and re-emitted luminosities is $\Omega = 1.4\pm0.2\pi$. However, this does not encompass the fluorescent yield of the gas, i.e. the probability that the creation of a vacancy in an inner shell is followed by radiative de-excitation rather than auto-ionization (i.e. the ejection of a second electron, known as the Auger effect). For Fe XXV and Fe XXVI no electron is available for the Auger mechanism, but the Ly$\alpha$ transition is the most likely result of the recombination cascade, so that the effective fluorescent yield is ~0.5 and 0.7, respectively (*52*). With such a correction, the solid angle of the wind is again $\Omega > 2\pi$.

## 2.2 Column density

The column density $N_H = 6.9^{+0.6}_{-1.2} \times 10^{23}$ cm$^{-2}$ of the highly ionized absorbing (and re-emitting) gas is a direct output of the spectral model in the broadband analysis. Since ion abundances slightly depend on the resolution of XSTAR grids, a lower sampling over the ionization parameter range might lead to underestimated opacities. The generation and implementation of finely sampled grids over a large area of the parameter space is rather time-consuming. Having identified the region the wind spectral components likely fall in, we produced another set of finer XSTAR tables with 15 linear steps at $5 < \log (\xi/\text{erg cm s}^{-1}) < 8$. With a negligible improvement in the overall fit statistic ($\Delta\chi^2 < -2$), the column density increases to $N_H = 8.8^{+0.4}_{-0.5} \times 10^{23}$ cm$^{-2}$. The values of $\log (\xi/\text{erg cm s}^{-1})$ grow by ~0.1, remaining consistent with those given in Table S3, and none of the other variables are affected.

Another estimate of $N_H$ is based on the optical depth $\tau_{tot}$ from the P-Cygni fit, which is related to the ionic column density of reprocessing gas $N_i$ through the expression $\tau_{tot} = (\pi e^2/m_e c) f \lambda_0 N_i/v_\infty$, where $e$ and $m_e$ are the electron charge (in statcoulomb) and mass, $f$ is the oscillator strength of a given transition, and $\lambda_0$ (in cm) is the laboratory wavelength of the line (*33*). For an identification with Fe XXVI Ly$\alpha$, which has a mean $f = 0.21$ for the resonant line doublet (*53*), the average $\tau_{tot} \sim 0.12$ for Obs. 2–5 implies $N_i \sim 1.3\pm0.2 \times 10^{19}$ cm$^{-2}$. At the measured ionization state of the absorbing gas, the Fe XXVI Ly$\alpha$ ionic fraction is close to 1 (*54*). Thus, for standard Solar iron abundance (*55*), this suggests an equivalent hydrogen column density $N_H = 4.0\pm0.6 \times 10^{23}$ cm$^{-2}$, comparable to that found through the XSTAR photo-ionization code.

## 2.3 Outflow velocity

The terminal velocity of the wind is one of the fit parameters in the custom P-Cygni model. The best value is $v_\infty = 0.35\pm0.02c$, and is consistent with the velocity corresponding to the full width at zero intensity of the absorption trough in the Gaussian-line fit. While this should be regarded as the actual speed of the outflowing gas, in keeping with the standard practice for narrow, unresolved lines, we adopt for the computation of the wind's energetics the more conservative line-of-sight bulk velocity $v_{out} = 0.25\pm0.01c$, averaged over the viewing angle through the absorbing gas and commensurate with the energy of maximum depth in the absorption profile.

## 2.4 Launch radius

The large uncertainty on the location of the absorbing gas usually prevents an accurate determination of the wind's energetics. The best way to overcome this limitation is through a timing approach, in order to measure how absorption varies over timescales of days/weeks. Since this is possible only for a small number of well-studied sources, a single-epoch estimate of the distance of the absorber can be derived from the spectral analysis, based on the ionization parameter of the gas $\xi = L_{ion}/nR^2$. The thickness $\Delta r$ of the gas layer responsible for the high-ionization absorption is lower than or equal to its maximum distance $R_{max}$ from the central source. Since $N_H \sim n\Delta r$, this condition becomes $R_{max} < L_{ion}/N_H\xi$. With an average $\xi \sim 10^6$ erg cm s$^{-1}$, and $L_{ion} \sim 5 \times 10^{46}$ erg/s (obtained from the SED fit), we get $R_{max} < 7 \times 10^{16}$ cm or $\sim 500\ r_g$. This is compatible with a recombination time of nearly one week for an inferred electron density of $\sim 10^7$ cm$^{-3}$.

The present campaign on PDS 456 allowed us to witness the considerable spectral variability of the source on relevant timescales to constrain $R_{in}$. The most dramatic variations occur from Obs. 1 to Obs. 2, when the hard X-ray power-law flux measured by *NuSTAR* decreases by a factor of three over the span of about ten days, and the Fe XXVI K$\alpha$ line flux also drops significantly (Table S2). In general, the wind emission follows the same variability pattern of the high-energy continuum in the five epochs, suggesting that the gas is in photo-ionization equilibrium. In fact, the variations of the ionization parameter $\xi$ correlate fairly well with those of the overall intensity. The light-crossing time between Obs. 1 and Obs. 2 corresponds to a distance of $\sim 2.6 \times 10^{16}$ cm (170 $r_g$), and sets an upper limit to $R_{in}$. Even if we cannot perform a proper reverberation study, $R_{in}$ is unlikely to be much smaller than this value. The order-of-magnitude estimate of a few hundreds of $r_g$ is further supported by the sizable changes in the depth of the absorption line over a week, observed just a few months earlier during the *Suzaku* monitoring (*18*).

In principle, $R_{in}$ can be as low as the escape radius, which is the innermost starting point of the wind to be physically acceptable and lies between $\sim 16\ r_g$ (for $v_\infty = 0.35c$) and 32 $r_g$ (for $v_{out} = 0.25c$). Even in this case, the corresponding mass loss would still be a sufficient fraction of the Eddington accretion rate to produce significant feedback (see below). However, such a small spatial extent would likely give rise to spectral variability on timescales of $\sim 1$–2 days (i.e. detectable within the single observations), for which no clear evidence was found.

3. Wind energetics

There is no direct measurement for the mass of the central black hole in PDS 456. An estimate can be obtained from the scaling relations between the virial radius of the broad-line region (BLR) and the black hole mass, based on reverberation studies, $M_{BH} = G^{-1} R_{BLR} V_{BLR}^2$, where $V_{BLR}$ is the Keplerian velocity of the BLR clouds. With $\lambda L_\lambda (5100\ \text{Å}) \sim 2 \times 10^{46}$ erg/s (*56*) and FWHM(H$\beta$) = 3974±764 km/s (*30*), we compute $R_{BLR}$ using the latest $R_{BLR}$–$\lambda L_\lambda(5100\ \text{Å})$ relationship (*57*), and assume $V_{BLR} \sim$ FWHM(H$\beta$) neglecting any correction factor of order unity to account for the BLR geometry. This results in log$(M_{BH}/M_{sun}) = 9.24 \pm 0.17$. Given the uncertainties, and keeping in mind that all the subordinate quantities below are accurate within a factor of $\sim 2$, we have assumed for simplicity a value of $M_{BH} = 10^9\ M_{sun}$ throughout this work. This implies for PDS 456 an Eddington luminosity $L_{Edd} \sim 1.3 \times 10^{47}$ erg/s, and an Eddington mass accretion rate $\dot{M}_{Edd} \sim 1.4 \times 10^{27}$ g/s or 22 $M_{sun}$/yr (for a radiative efficiency $\eta \sim 0.1$).

Based on the discussion in the previous section, we adopt $\Omega = 2\pi$, $N_H = 6 \times 10^{23}$ cm$^{-2}$, $v_{out} = 0.25 \pm 0.01 c$, and $R_{in} = 1.5 \times 10^{16}$ cm in Eq. 4, and calculate a mass outflow rate:

$$\dot{M}_{out} \sim 7 \times 10^{26} \text{ g/s} \sim 11\ M_{sun}/\text{yr} \sim 0.5\ \dot{M}_{Edd}. \qquad (6)$$

Under the same assumptions, the kinetic power of the wind is therefore:

$$P_{kin} = \tfrac{1}{2} \dot{M} v_{out}^2 \sim 2 \times 10^{46} \text{ erg/s} \sim 0.15\ L_{Edd}. \qquad (7)$$

While the above fractions of Eddington limits depend on the actual black hole mass, the absolute values of the mass loss and of the associated kinetic power do not. We can then determine the total mechanical power $E_{wind}$ released by the AGN through the wind over a typical lifetime of $10^8$ yr (*58*). We take a duty cycle of 0.1. In other words, the wind is in action for ~$10^7$ yr. The fraction of AGN with detected X-ray signatures of fast outflows (about 40%) can be interpreted in two different ways: for a duty cycle of 1, this is a measure of the covering fraction $\Omega/4\pi$, whereas it would indicate the wind lifetime in case of full covering. Unless the wide-angle, high-column, high-velocity phase of the wind is extremely short-lived, our guess is therefore rather conservative, and leads to $E_{wind} \sim 6 \times 10^{60}$ erg. In order to understand the possible impact in terms of feedback, this has to be compared to the binding energy of the galactic bulge in a system like PDS 456, whose order of magnitude has been again evaluated from the scaling relations as $E_{grav} \sim M_{bulge}\,\sigma_*^2$, where $M_{bulge}$ is the dynamical mass of the bulge and $\sigma_*$ is the stellar velocity dispersion. For a $10^9\,M_{sun}$ black hole, $M_{bulge} \sim 10^{12}\,M_{sun}$ (*59, 60*), while $\sigma_* \sim 330$ km/s (*61, 62*). Hence $E_{grav} \sim 2 \times 10^{60}$ erg, and the condition $E_{wind} \gtrsim E_{grav}$ safely holds for log $(M_{BH}/M_{sun}) < 9.5$. Since the deposition of a few per cent of the wind power in the interstellar matter can be enough to trigger a snowplow effect on galactic scales, even the constraints on both the actual launch radius (which can be closer to the escape radius if the matter is too ionized to produce any detectable effect) and the wind lifetime can be significantly relaxed. Indeed, even a modest coupling (with ~10% efficiency) between the disk wind and any energy-conserving, large-scale outflow would culminate in the usual values of mass-loss rate (~$10^3\,M_{sun}$/yr) and velocity (~$10^3$ km/s) observed in the molecular gas component of several local AGN (*22*).

With respect to the wind driving mechanism in PDS 456, no persistent outflow can arise from resonant line absorption due to the extreme ionization state of the gas. For sources accreting at about the Eddington limit, however, the wind is expected to be optically thick to electron scattering, so that continuum radiation pressure becomes effective (*63*). Here the best-fit column density of the highly ionized absorber implies a Thomson optical depth of $\tau \sim 0.5$, which is in practice a lower limit, since additional, fully ionized gas would remain undetected. Assuming $\tau = 1$, conservation of momentum requires that $\dot{M}_{out}v_{out} \sim L_{bol}/c$. For the derived wind parameters, we find that the momentum of the gas exceeds the momentum of the photons by a factor of ~1.5. Given the uncertainties on $\dot{M}_{out}$ and $L_{bol}$, this is a reasonable agreement with momentum conservation. Moreover, it is possible that some other form of driving (e.g. magnetic) is also present, further alleviating any tension.

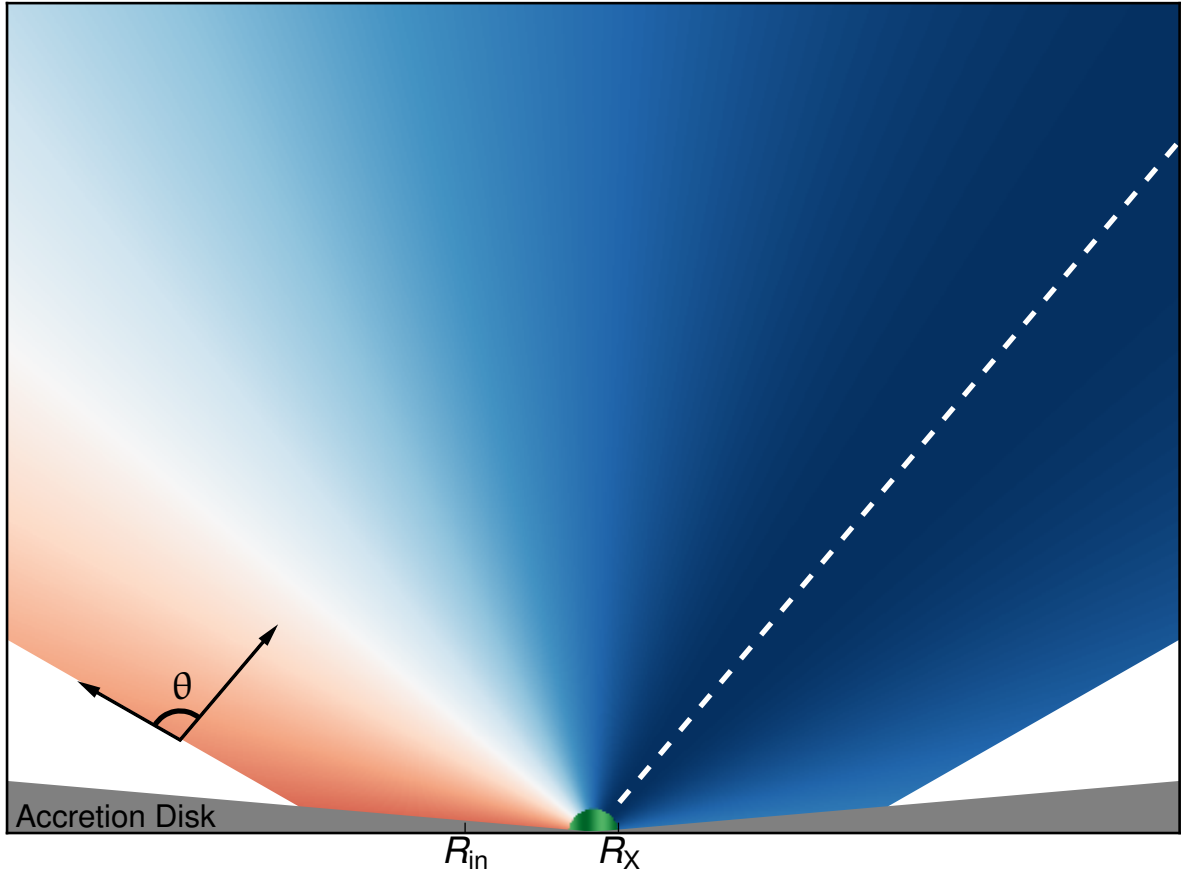

**Fig. S1.** For a preliminary, qualitative evaluation of the wind's opening angle θ, we can consider a radial outflow with uniform velocity. Due to the Doppler effect, the different energies in the P-Cygni profile correspond to different velocity components projected along the line of sight (marked by the dashed line). In this diagram (not to scale), the angular color code indicates the degree of blueshift/redshift of the Fe XXVI Kα line for various directions within the outflow. In this oversimplified framework, the observed range of velocities implies that $\theta \sim \cos^{-1}(v_{min}/v_{max}) \sim 100$ degrees (hence $\Omega \sim 3\pi$ sr). It is worth noting that the overall geometrical structure (e.g. polar versus equatorial) of the outflow is actually unknown, and no assumptions have been made in the derivation of its solid angle. This notwithstanding, the flux variability timescale in PDS 456 suggests a size of the X-ray source of the order of $R_X \sim 10\ r_g$, while the starting point of the wind is most likely around $R_{in} \sim 100\ r_g$.

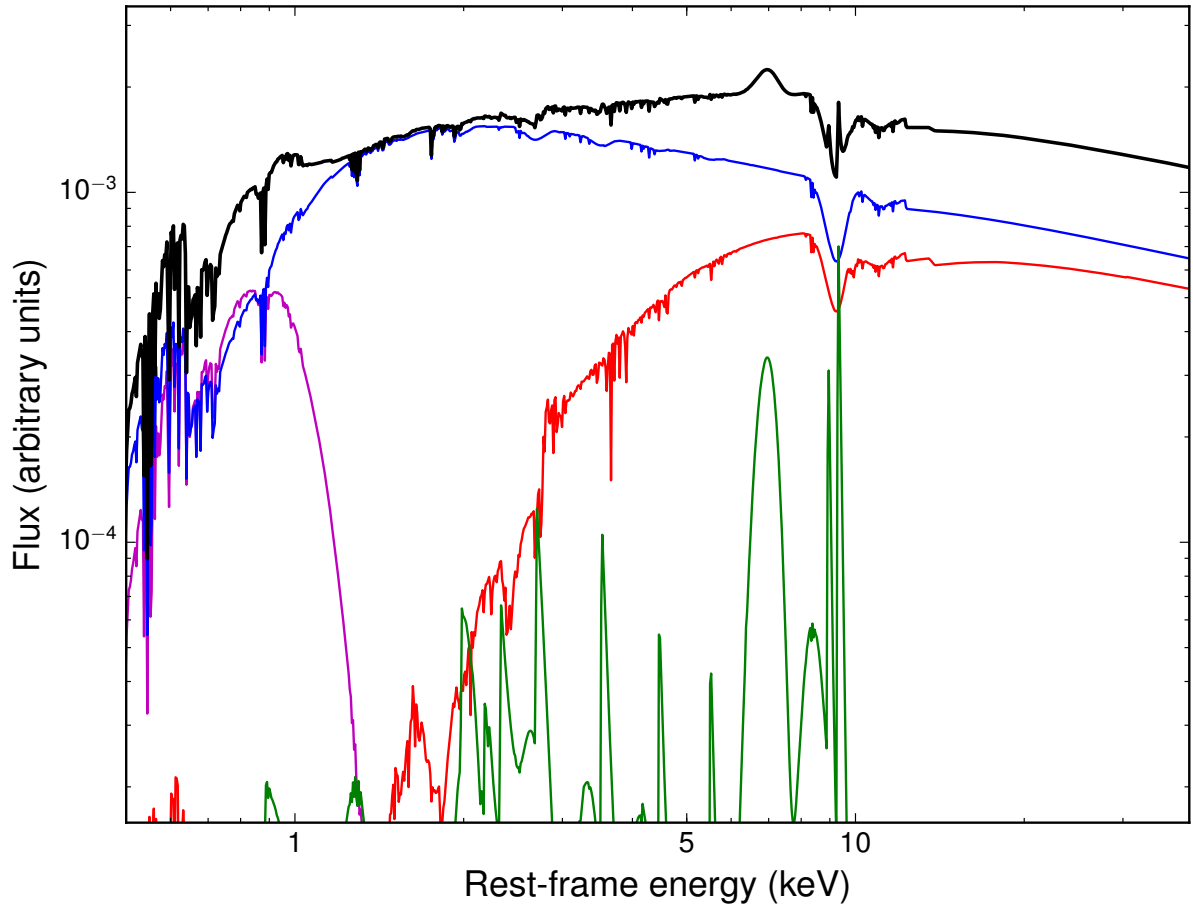

**Fig. S2.** Template of the self-consistent broadband model (black curve; based on Obs. 4) with all the emission components disentangled: primary power-law continuum, either direct (blue) or transmitted through the partial covering layer (red), wind emission (green) and soft excess (magenta). The absorption features imprinted by the highly ionized gas are clearly appreciable. Note, however, that CCD detectors have a resolving power of only ~20–50, for an energy resolution of around 80 eV at 1 keV, and 150 eV at 6 keV.

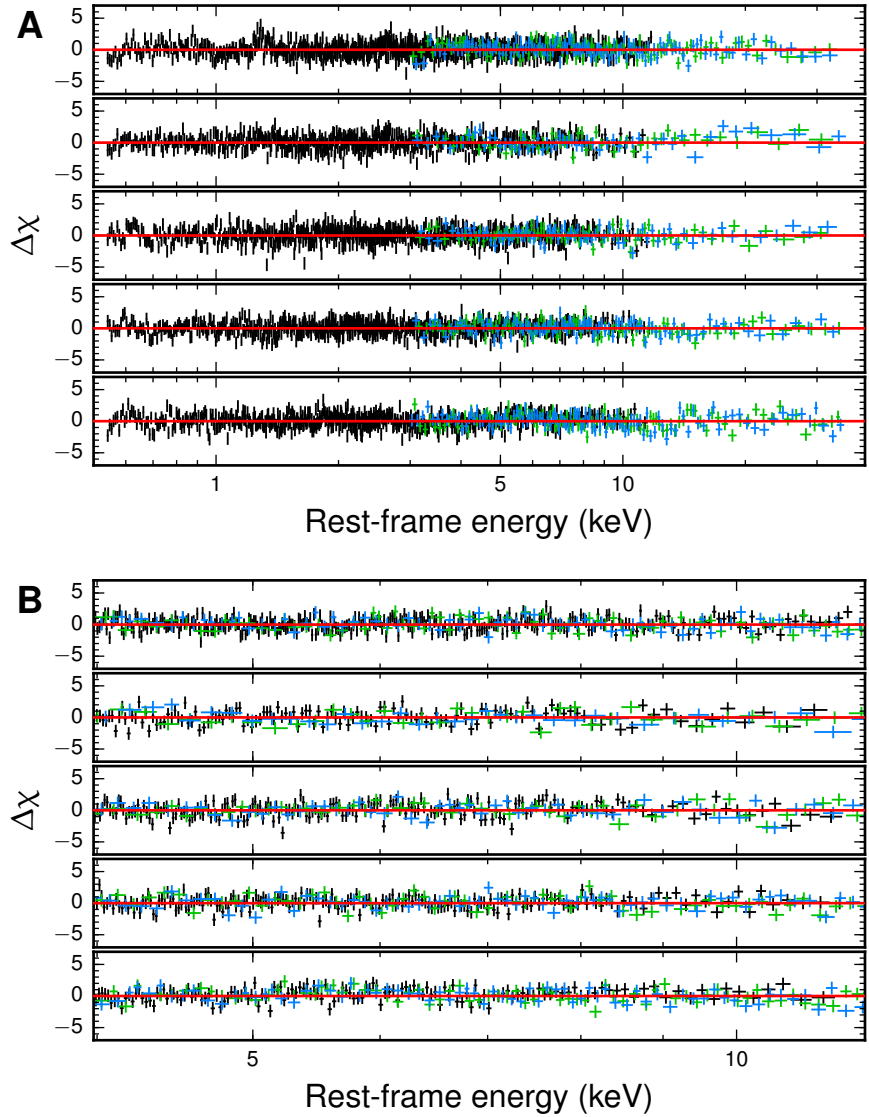

**Fig. S3.** (A) Residuals with respect to the best-fitting wind model ($\chi^2/\nu = 4372.5/4180$) for each of the five observations 1–5 (in descending order), plotted in units of SD with error bars of size one. (B) Zoom on the 4–12 keV energy range (rest-frame), showing that no emission/absorption structure is left unmodeled in the Fe-K band ($\chi^2/\nu = 1465/1460$).

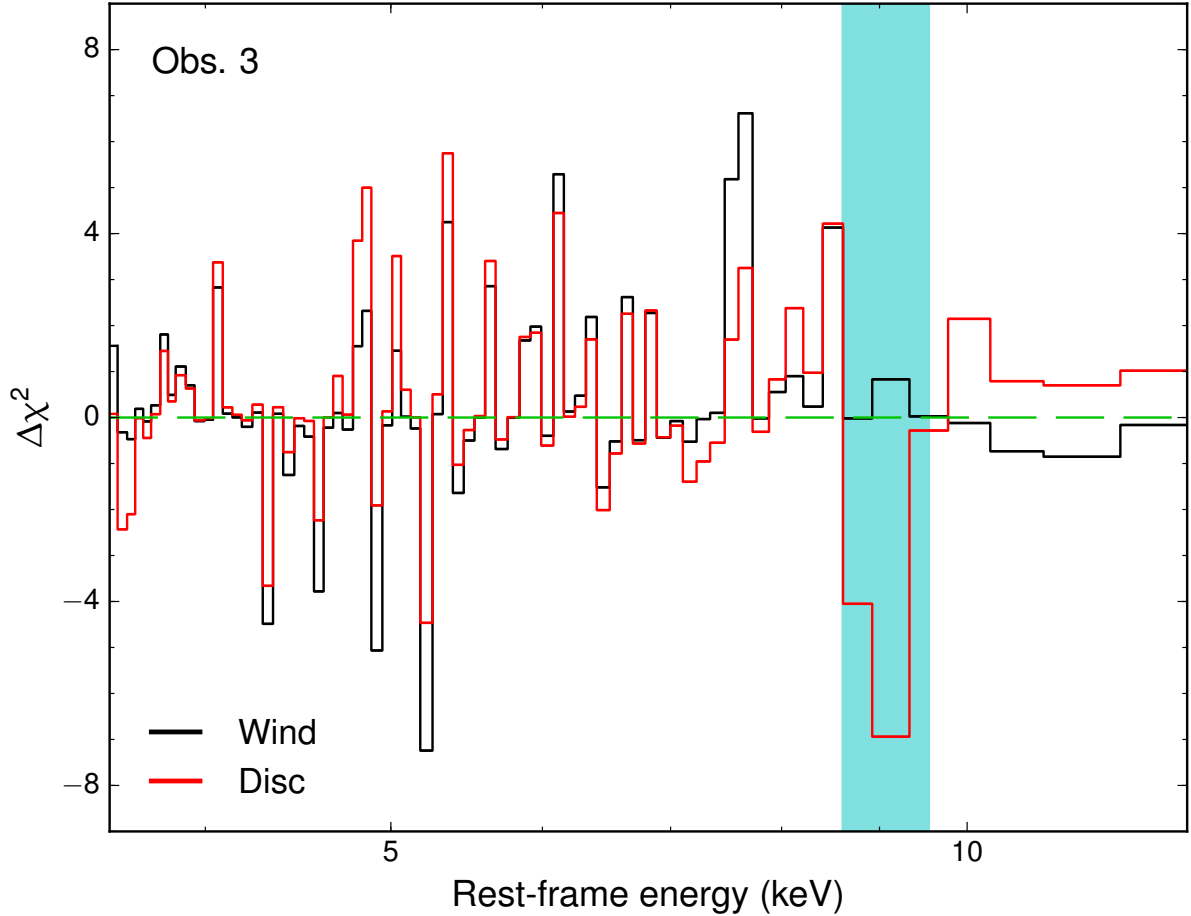

**Fig. S4.** Residuals over the best fit for Obs. 3 for both the disk absorption (in red) and wind absorption (in black) models. The former scenario is not capable of accounting for the depth of the absorption trough around 9 keV even if 10× Solar Fe abundance is adopted. This is due to the fact that the observed continuum is not dominated by a disk reflection component (see the text for thorough discussion). The shaded area indicates the energy range corresponding to the full width at half maximum of the absorption feature when fitted with a simple Gaussian profile. Only *XMM–Newton* data (re-binned) are plotted for clarity.

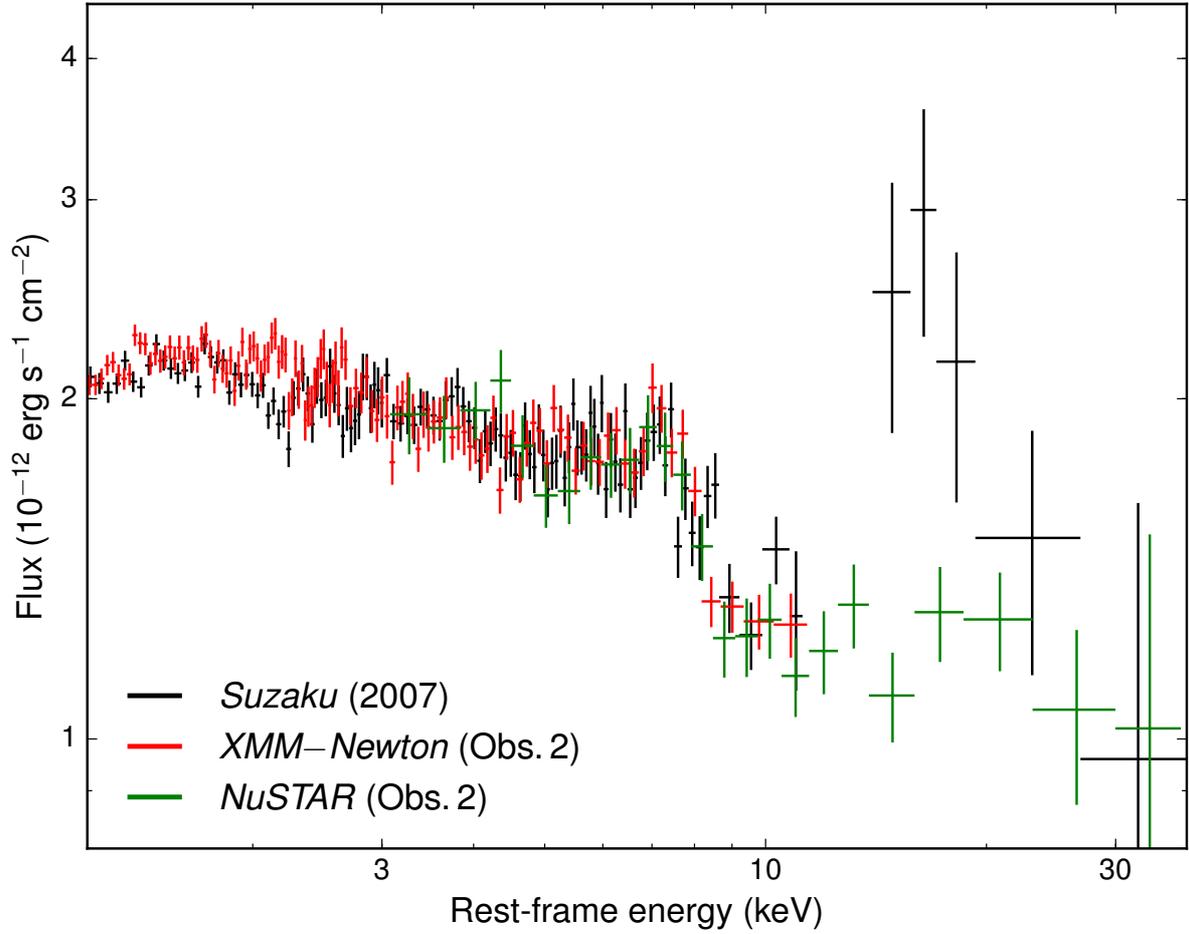

**Fig. S5.** Comparison between Obs. 2 of the current *XMM–Newton* and *NuSTAR* campaign and the 2007 *Suzaku* observation of PDS 456 (the latter scaled down by a factor of ~0.75). While the hard excess tentatively detected by *Suzaku* is consistent with a standard (RS ~ 1) contribution from disk reflection to the observed spectrum, a similar Compton hump is not present in any of the much more sensitive *NuSTAR* data sets. This implies that any reflection is quite faint and hard to constrain, possibly due to a different configuration of the X-ray illuminating source.

**Table S1.** *XMM–Newton* and *NuSTAR* observation log. $T_{\rm tot}$: total elapsed time. $T_{\rm net}$: net exposure after standard screening. Background subtracted count rates refer to the 0.45–10 keV (*XMM–Newton*) and 2.5–30 keV (*NuSTAR*) bands.

| Obs. | Satellite | Obs. Start (UT) | Obs. End (UT) | $T_{\rm tot}$ (ks) | Detector | $T_{\rm net}$ (ks) | Count rate (s$^{-1}$) |
|---|---|---|---|---|---|---|---|
| 1 | *XMM–Newton* | 2013-08-27, 04:41 | 2013-08-28, 11:13 | 110.0 | pn | 95.7 | $3.083 \pm 0.006$ |
|   |              |                   |                   |       | MOS1 | 105.8 | $0.730 \pm 0.003$ |
|   |              |                   |                   |       | MOS2 | 107.7 | $0.932 \pm 0.003$ |
|   | *NuSTAR*     | 2013-08-27, 03:41 | 2013-08-28, 11:41 | 113.9 | FPMA | 43.8 | $0.165 \pm 0.002$ |
|   |              |                   |                   |       | FPMB | 43.8 | $0.156 \pm 0.002$ |
| 2 | *XMM–Newton* | 2013-09-06, 03:24 | 2013-09-07, 10:36 | 112.3 | pn | 92.2 | $1.990 \pm 0.005$ |
|   |              |                   |                   |       | MOS1 | 106.3 | $0.545 \pm 0.002$ |
|   |              |                   |                   |       | MOS2 | 106.3 | $0.566 \pm 0.002$ |
|   | *NuSTAR*     | 2013-09-06, 02:56 | 2013-09-07, 10:51 | 113.9 | FPMA | 42.9 | $0.064 \pm 0.001$ |
|   |              |                   |                   |       | FPMB | 43.0 | $0.061 \pm 0.001$ |
| 3 | *XMM–Newton* | 2013-09-15, 18:47 | 2013-09-17, 03:57 | 119.4 | pn | 102.0 | $1.883 \pm 0.004$ |
|   |              |                   |                   |       | MOS1 | 111.0 | $0.501 \pm 0.002$ |
|   |              |                   |                   |       | MOS2 | 111.4 | $0.574 \pm 0.002$ |
|   | *NuSTAR*     | 2013-09-15, 17:56 | 2013-09-17, 04:01 | 119.1 | FPMA | 44.0 | $0.089 \pm 0.002$ |
|   |              |                   |                   |       | FPMB | 44.0 | $0.083 \pm 0.002$ |
| 4 | *XMM–Newton* | 2013-09-20, 02:47 | 2013-09-21, 09:37 | 111.0 | pn | 93.0 | $1.854 \pm 0.004$ |
|   |              |                   |                   |       | MOS1 | 107.8 | $0.519 \pm 0.002$ |
|   |              |                   |                   |       | MOS2 | 107.7 | $0.574 \pm 0.002$ |
|   | *NuSTAR*     | 2013-09-20, 03:06 | 2013-09-21, 11:11 | 113.8 | FPMA | 58.5 | $0.095 \pm 0.001$ |
|   |              |                   |                   |       | FPMB | 58.6 | $0.089 \pm 0.001$ |
| 5 | *XMM–Newton* | 2014-02-26, 08:03 | 2014-02-27, 22:51 | 139.7 | pn | 100.3 | $1.323 \pm 0.004$ |
|   |              |                   |                   |       | MOS1 | 119.1 | $0.415 \pm 0.002$ |
|   |              |                   |                   |       | MOS2 | 110.7 | $0.411 \pm 0.002$ |
|   | *NuSTAR*     | 2014-02-26, 08:16 | 2014-02-28, 22:56 | 224.3 | FPMA | 109.7 | $0.060 \pm 0.001$ |
|   |              |                   |                   |       | FPMB | 109.3 | $0.059 \pm 0.001$ |

**Table S2.** Preliminary parameterization of the Fe XXVI K-shell features. $\Gamma$: photon index. $f_{cov}$: covering fraction. $\sigma$: Gaussian line width. $\tau_{max}$: optical depth of the absorption edge. $\Delta\chi^2$: change in the overall statistics (for an extra degree of freedom) after feature removal and re-fitting. $E_c$: characteristic energy of the P-Cygni profile (in keV). $v_\infty$: terminal outflow velocity. $\tau_{tot}$ and $\alpha_{1,2}$ control the dependence of the optical depth of the reprocessing gas on the velocity field (*10, 32*). (f) and (t) are frozen and tied parameters, respectively. Energies are given in the rest frame, while continuum fluxes are intrinsic at 2–30 keV. All errors correspond to 90% confidence levels for the single parameters.

| Component Parameter | Observation 1 | 2 | 3 | 4 | 5 |
|---|---|---|---|---|---|
| **Continuum** | | | | | |
| *Power law* | | | | | |
| $\Gamma$ | $2.29^{+0.04}_{-0.03}$ | (t) | (t) | (t) | (t) |
| $F\,(10^{-12}\,\mathrm{erg\,s^{-1}\,cm^{-2}})$ | $11.6^{+0.9}_{-0.8}$ | $4.3^{+0.4}_{-0.2}$ | $6.7^{+0.5}_{-0.4}$ | $7.3^{+0.6}_{-0.5}$ | $4.8^{+0.4}_{-0.3}$ |
| *Partial covering* | | | | | |
| $N_{\mathrm{H}}\,(10^{22}\,\mathrm{cm^{-2}})$ | $12\pm1$ | (t) | (t) | (t) | (t) |
| $f_{\mathrm{cov}}$ | $0.32^{+0.04}_{-0.03}$ | $0.06^{+0.06}_{-0.05}$ | $0.39^{+0.02}_{-0.04}$ | $0.40^{+0.02}_{-0.03}$ | $0.33\pm0.04$ |
| **Fe K emission** | | | | | |
| *Fe XXVI Ly$\alpha$* | | | | | |
| Energy (keV) | 6.97(f) | (t) | (t) | (t) | (t) |
| $\sigma_K$ (eV) | $380^{+50}_{-30}$ | (t) | (t) | (t) | (t) |
| $F\,(10^{-14}\,\mathrm{erg\,s^{-1}\,cm^{-2}})$ | $7.2\pm0.3$ | $4.0\pm1.3$ | $2.6\pm1.6$ | $5.1\pm1.8$ | $1.5\pm1.3$ |
| EW (eV) | $100\pm30$ | $145\pm100$ | $<140$ | $110\pm80$ | $<135$ |
| $v_{\mathrm{out}}/c$ | $0.029\pm0.017$ | (t) | (t) | (t) | (t) |
| **Fe K absorption** | | | | | |
| *Fe XXVI Ly$\alpha$* | | | | | |
| Energy (keV) | 6.97(f) | (t) | (t) | (t) | (t) |
| $\sigma_K$ (eV) | 380(t) | (t) | (t) | (t) | (t) |
| $F\,(10^{-14}\,\mathrm{erg\,s^{-1}\,cm^{-2}})$ | $-4.3^{+2.1}_{-2.5}$ | $-5.2^{+1.2}_{-1.5}$ | $-11.1^{+1.5}_{-1.9}$ | $-9.7^{+1.5}_{-2.1}$ | $-7.9^{+1.2}_{-1.5}$ |
| EW (eV) | $-75\pm45$ | $-250\pm70$ | $-350\pm60$ | $-280\pm50$ | $-345\pm55$ |
| $v_{\mathrm{out}}/c$ | $0.239^{+0.004}_{-0.005}$ | (t) | (t) | (t) | (t) |
| *Fe XXVI Ly$\beta$* | | | | | |
| Energy (keV) | 8.25(f) | (t) | (t) | (t) | (t) |
| $\sigma_K$ (eV) | 380(t) | (t) | (t) | (t) | (t) |
| $F\,(10^{-14}\,\mathrm{erg\,s^{-1}\,cm^{-2}})$ | $>-3.9$ | $-4.1\pm1.5$ | $-7.4\pm1.8$ | $-4.3^{+1.8}_{-2.0}$ | $-5.1^{+1.2}_{-1.4}$ |
| EW (eV) | $>-90$ | $-250\pm90$ | $-290\pm70$ | $-160\pm70$ | $-280\pm70$ |
| $v_{\mathrm{out}}/c$ | 0.239(t) | (t) | (t) | (t) | (t) |
| $\Delta\chi^2$ | 0.7 | 20.5 | 44.1 | 14.5 | 43.6 |
| *Fe XXVI K edge* | | | | | |
| Energy (keV) | 9.28(f) | (t) | (t) | (t) | (t) |
| $\tau_{\max}$ | $<0.06$ | $0.18^{+0.11}_{-0.14}$ | $0.31^{+0.12}_{-0.11}$ | $0.29\pm0.10$ | $0.41^{+0.11}_{-0.09}$ |
| $v_{\mathrm{out}}/c$ | 0.239(t) | (t) | (t) | (t) | (t) |
| $\Delta\chi^2$ | – | 4.3 | 21.1 | 23.6 | 52.9 |
| **P-Cygni profile** | | | | | |
| $E_c$ (keV) | $7.42^{+0.09}_{-0.08}$ | (t) | (t) | (t) | (t) |
| $v_\infty/c$ | $0.35\pm0.02$ | (t) | (t) | (t) | (t) |
| $\tau_{\mathrm{tot}}$ | $0.05\pm0.01$ | $0.12\pm0.03$ | $0.13\pm0.02$ | $0.11\pm0.02$ | $0.12\pm0.02$ |
| $\alpha_{1,2}$ | $6.4^{+2.4}_{-1.6}$ | (t) | (t) | (t) | (t) |

**Table S3.** Best-fit parameters of the broadband wind model. $v_K$: velocity smearing of the photo-ionization spectral grids. $\xi$: ionization parameter of the gas. $\kappa_{\rm obs}$: normalization of the wind emission component. $F_K$: intrinsic ionizing flux for Fe XXVI (7–30 keV rest-frame). All the other symbols are the same as defined in the legend of Table S2, except for the continuum and soft excess fluxes that now refer to the 0.45–30 keV band.

| Component Parameter | Observation 1 | 2 | 3 | 4 | 5 |
|---|---|---|---|---|---|
| **Galactic absorption** | | | | | |
| $N_{\rm H,Gal}$ ($10^{22}$ cm$^{-2}$) | 0.2(f) | (t) | (t) | (t) | (t) |
| **Primary continuum** | | | | | |
| $\Gamma$ | $2.28^{+0.01}_{-0.02}$ | $2.68 \pm 0.02$ | $2.43 \pm 0.02$ | $2.35 \pm 0.02$ | $2.39^{+0.01}_{-0.02}$ |
| $F$ ($10^{-12}$ erg s$^{-1}$ cm$^{-2}$) | $25 \pm 3$ | $16 \pm 2$ | $16 \pm 2$ | $16 \pm 2$ | $10 \pm 1$ |
| **Wind absorption** ($v_K = 15,000$ km/s) | | | | | |
| $N_{\rm H}$ ($10^{22}$ cm$^{-2}$) | $69^{+6}_{-12}$ | (t) | (t) | (t) | (t) |
| $\log(\xi/{\rm erg\,cm\,s^{-1}})$ | $6.30^{+0.10}_{-0.16}$ | $5.99^{+0.14}_{-0.13}$ | $5.75^{+0.06}_{-0.16}$ | $5.80^{+0.07}_{-0.15}$ | $5.84^{+0.07}_{-0.12}$ |
| $v_{\rm out}/c$ | $0.254^{+0.009}_{-0.008}$ | $0.244^{+0.010}_{-0.014}$ | $0.251^{+0.004}_{-0.003}$ | $0.254^{+0.006}_{-0.003}$ | $0.240^{+0.006}_{-0.005}$ |
| **Wind emission** ($v_K = 15,000$ km/s) | | | | | |
| $N_{\rm H}$ ($10^{22}$ cm$^{-2}$) | 69(t) | (t) | (t) | (t) | (t) |
| $\log(\xi/{\rm erg\,cm\,s^{-1}})$ | 6.30(t) | 5.99(t) | 5.75(t) | 5.80(t) | 5.84(t) |
| $\kappa_{\rm obs}$ ($\times 10^{-3}$) | $1.47^{+0.58}_{-0.60}$ | $0.57^{+0.25}_{-0.24}$ | $0.38 \pm 0.16$ | $0.82^{+0.16}_{-0.20}$ | $0.59^{+0.16}_{-0.14}$ |
| $v_{\rm out}/c$ | $0.012 \pm 0.010$ | (t) | (t) | (t) | (t) |
| **Partial covering** ($v_K = 15,000$ km/s) | | | | | |
| $N_{\rm H}$ ($10^{22}$ cm$^{-2}$) | $25^{+2}_{-3}$ | $25^{+2}_{-3}$ | $18^{+2}_{-1}$ | $17 \pm 1$ | $6.8 \pm 0.4$ |
| $\log(\xi/{\rm erg\,cm\,s^{-1}})$ | $2.91 \pm 0.02$ | (t) | (t) | (t) | (t) |
| $f_{\rm cov}$ | $0.33^{+0.04}_{-0.05}$ | $0.44 \pm 0.06$ | $0.48 \pm 0.05$ | $0.45^{+0.05}_{-0.06}$ | $0.50 \pm 0.03$ |
| $v_{\rm out}/c$ | 0.254(t) | 0.244(t) | 0.251(t) | 0.254(t) | 0.240(t) |
| **Soft excess** | | | | | |
| $E_{\rm s}$ (keV) | $0.63 \pm 0.01$ | (t) | (t) | (t) | $0.80 \pm 0.01$ |
| $\sigma_{\rm s}$ (eV) | $196^{+6}_{-3}$ | (t) | (t) | (t) | $105^{+9}_{-10}$ |
| $F_{\rm s}$ ($10^{-13}$ erg s$^{-1}$ cm$^{-2}$) | $22^{+3}_{-1}$ | $23^{+3}_{-1}$ | $27^{+4}_{-2}$ | $22^{+3}_{-1}$ | $4.2 \pm 0.3$ |
| **Warm absorption** ($v_K = 100$ km/s) | | | | | |
| $N_{\rm H}$ ($10^{22}$ cm$^{-2}$) | $0.43^{+0.03}_{-0.01}$ | (t) | (t) | (t) | $0.18^{+0.01}_{-0.02}$ |
| $\log(\xi/{\rm erg\,cm\,s^{-1}})$ | $0.31 \pm 0.02$ | (t) | (t) | (t) | (t) |
| $v_{\rm out}/c$ | 0(f) | (t) | (t) | (t) | (t) |
| **Cross norm.** | | | | | |
| FPMA/pn | $1.03 \pm 0.02$ | $1.05^{+0.04}_{-0.03}$ | $1.07^{+0.04}_{-0.03}$ | $1.02^{+0.02}_{-0.03}$ | $1.01^{+0.03}_{-0.02}$ |
| FPMB/pn | $1.02^{+0.02}_{-0.03}$ | $1.05 \pm 0.04$ | $1.05 \pm 0.03$ | $1.00^{+0.02}_{-0.03}$ | $1.08^{+0.02}_{-0.03}$ |
| **7–30 keV flux** | | | | | |
| $F_K$ ($10^{-12}$ erg s$^{-1}$ cm$^{-2}$) | 5.6 | 1.8 | 2.8 | 3.2 | 1.9 |